\documentclass[aps,prd,showpacs,twocolumn,
superscriptaddress]{revtex4-1}

\usepackage{graphicx}
\usepackage{subfigure} 
\usepackage{dcolumn}

\graphicspath{{figures/}}
\usepackage{amsmath}
\usepackage{amssymb}
\usepackage{bm}
\usepackage{color}
\usepackage{comment}

\usepackage[normalem]{ulem}
\usepackage[dvipsnames]{xcolor}
\usepackage{hyperref}

\usepackage{verbatim} 

\hypersetup{
colorlinks=true, 
linkcolor=blue,  
citecolor=cyan,  
}

\newcommand{\bea}{\begin{eqnarray}}
\newcommand{\eea}{\end{eqnarray}}
\newcommand{\be}{\begin{equation}}
\newcommand{\ee}{\end{equation}}

\begin{document}
\title{Tests of no-hair theorem with binary black-hole coalescences}
\author{Song Li}
\email{leesong@shao.ac.cn}
\affiliation{Shanghai Astronomical Observatory, Shanghai, 200030, China}
\affiliation{School of Astronomy and Space Science, University of Chinese Academy of Sciences,
Beijing, 100049, China}

\author{Wen-Biao Han}
\email{wbhan@shao.ac.cn}
\affiliation{School of Fundamental Physics and Mathematical Sciences, Hangzhou Institute for Advanced Study, UCAS, Hangzhou 310024, China}
\affiliation{Shanghai Astronomical Observatory, Shanghai, 200030, China}
\affiliation{School of Astronomy and Space Science, University of Chinese Academy of Sciences,
Beijing, 100049, China}
\affiliation{International Centre for Theoretical Physics Asia-Pacific, Beijing/Hangzhou 100190, China}

\author{Shu-Cheng Yang}
\email{ysc@shao.ac.cn}
\affiliation{Shanghai Astronomical Observatory, Shanghai, 200030, China}

\date{\today}

\begin{abstract}
Test of the no-hair theorem is the primary target with gravitational waves from binary black holes. In this Letter, we analyze gravitational-wave data from the LIGO-Virgo-KAGRA detection of binary black-hole mergers using the $\Psi_{\mathrm{FD}}$ model, which is a non-general relativity full waveform template for arbitrary axisymmetric black holes. By analyzing two high signal-noise-ratio events, GW150914 and GW200129, the no-hair theorem is tested at a significance level of 95\%, which is the best constraint until now. Especially, we find a significant deviation from the Kerr black hole in GW200129. More events and further analysis are needed to validate this deviation. 
\end{abstract}

\maketitle

\textit{Introduction.}---In the past decade, significant advancements have been made in the study of gravitational waves (GWs). A major breakthrough occurred in 2015 when the LIGO-Virgo Collaboration detected GW150914, marking the first observation of gravitational wave events from a binary black hole system\cite{GW150914}. Subsequently, the LIGO-Virgo-KAGRA Collaboration has reported the discovery of more compact binary systems. Over 90 of these systems have been identified to date, including two neutron star and black hole systems that are distinct from other binary black hole systems\cite{GW_events_1, GW_events_2, GW_events_3, GW_events_4}. These unique compact binaries present excellent opportunities to test general relativity (GR)\cite{Test_GR_1, Test_GR_2, Test_GR_3, Test_GR_4, Test_GR_5, Test_GR_6} and gain new insights into compact objects\cite{CO_1}, potentially leading to the discovery of new theories beyond GR. Ground-based detectors such as LIGO, Virgo, and KAGRA have been crucial in advancing our understanding of compact binary physics and astrophysics. Looking ahead, future space-based detectors like the Laser Interferometer Space Antenna (LISA), Taiji\cite{taiji}, and Tianqin\cite{tianqin} will provide new opportunities to enhance our comprehension of the Universe further.

According to the no-hair theorem, black holes in general relativity can be identified by just a few key characteristics: their mass, electric charge, and angular momentum. Other attributes, such as the matter that formed the black hole, do not affect its external gravitational field. The coalescence of two black holes consists of three stages: inspiral, merger, and ringdown. During the inspiral phase, the black holes gradually approach each other before the merger occurs. At the merger stage, they combine and form a single black hole. Finally, in the ringdown phase, the newly-formed black hole undergoes oscillations, gradually releasing excess energy and angular momentum in the form of gravitational waves until it reaches a stable state of equilibrium. Different methods are used to describe each phase. The inspiral phase is typically described using post-Newtonian theory, which provides an accurate representation of the gradual approach and orbital decay of the black holes. On the other hand, the merger phase is best described by numerical relativity simulations, which capture the complex dynamics and final coalescence of the black holes. To study the ringdown phase, where there is a slight perturbation in spacetime compared to a stable black hole, quasinormal modes(QNMs) are employed. These modes provide insights into the characteristic oscillations and decay of the gravitational waves. Additionally, black-hole perturbation theory is utilized to study the ringdown waveform, accounting for these perturbations.
Well-known waveform models, such as EOBNR and IMRPhenom\cite{Model_Nor_1, Model_Nor_2, Model_Nor_3, Model_Nor_4, Model_Nor_5, Model_Nor_6, Model_Nor_7, Model_Nor_8, Model_Nor_9, Model_Nor_10}, rely on these methods to accurately predict and analyze the gravitational wave signatures during black hole coalescence.

The Event Horizon Telescope (EHT) collaboration has recently achieved a remarkable feat by capturing shadow images of collapsed objects situated at the centers of two galaxies: M87 in an elliptical galaxy and Sagittarius-A* (SgrA*) in the Milky Way\cite{EHT_1, EHT_2, EHT_3, EHT_S}. These groundbreaking observations have opened up a new avenue of research dedicated to testing gravity theories and black hole solutions within a previously unexplored gravitational field regime\cite{Test_GR_Sha_1, Test_GR_Sha_2, Test_GR_Sha_3}. The appearance of these images is the consequence of light bending in the gravitational field of the source, a phenomenon that has been extensively studied within the framework of general relativity (GR)\cite{Light_Bend_1, Light_Bend_2, Light_Bend_3, Light_Bend_4}. Furthermore, researchers have utilized the shadow image of M87 to investigate alternative theories of gravity, including superspinar\cite {Test_GR_Sha_3} and conformal massive gravity\cite {Jusufi_20}. In addition, McWilliams\cite{BOB} proposed a novel waveform model for GR black holes called the backwards one-body (BOB) method, which solely considers photon motion without any phenomenological degrees of freedom.

Various methods have been employed to test the no-hair theorem using different instruments, including ground-based gravitational wave detectors and EHT. Isi\cite{Test_GR_2} specifically focused on analyzing the event GW150914 and evaluated the applicability of the no-hair theorem by studying the ringdown waveform. They claimed to have tested the no-hair theorem at approximately 10\%. Broderick\cite{Broderick_2014} employed a quasi-Kerr metric, which introduces an independent quadrupole moment, to examine the validity of the theorem through a simulated image of SgrA*. While they presented the first simulated images of a radiatively inefficient accretion flow (RIAF) around SgrA* using a quasi-Kerr metric, the results show weak constraints on potential modifications of the Kerr metric. Motivated by these articles, we have chosen a non-general relativity full waveform template called $\Psi_{\mathrm{FD}}$ to analyze different binary black holes observed during the public LIGO-Virgo-KAGRA (LVK) data. By analyzing different events, we aim to identify deviations in the quadrupole moment from that of a Kerr black hole to test the no-hair theorem.


\textit{Waveform Model.}---The gravitational waveform model employed in this letter is the $\Psi_{\mathrm{FD}}$ model\cite{Li_23}, a non-general relativity full waveform template for arbitrary axisymmetric black holes. Because the $\Psi_{\mathrm{FD}}$ model is derived from the IMRPhenomPv2 model, it can be used to describe the gravitational-wave signal of nonprecessing black-hole binaries.

The phase ansatz in the inspiral stage is given by:

\begin{equation}
\begin{aligned}
\phi_{\mathrm{Ins}}= & \phi_{\mathrm{TF} 2}(M f ; \Xi) \\
& +\frac{1}{\eta}\left(\sigma_{0}+\sigma_{1} f+\frac{3}{4} \sigma_{2} f^{4 / 3}+\frac{3}{5} \sigma_{3} f^{5 / 3}+\frac{1}{2} \sigma_{4} f^{2}\right)\\
&+\phi_{\Delta} \label{phase_inspiral}
\end{aligned}
\end{equation}

where the $\phi_{\mathrm{TF} 2}$ is the full TaylorF2 phase, the constants $\sigma_{i}$ (where $i = 0, 1, 2, 3, 4$) denote the correlation between the mass and spin of the system. Additionally, the phase deformation caused by a general parametrized black hole, different from the Kerr black hole, is represented as $\phi_{\Delta}$, which can be found in Ref.~\cite{Li_23}. 


The following equation gives the quadrupole moment in the general parametrized black hole:
\begin{equation}
{Q}=-Ma^2-B_2M^3\sqrt{5/4\pi}={Q}_K+\Delta{Q} \label{Total_Q}
\end{equation}
where $B_2$ is a dimensionless parameter associated with the phase deformation $\phi_{\Delta}$, $Q_K$ signifies the quadrupole moment of the Kerr black hole, and $\Delta Q$ illustrates the deviation in the quadrupole moment from the parametrized black hole. If we can estimate the value of $\Delta Q$ based on various gravitational events, we can then examine the validity of general relativity (GR) at that particular level. This is our objective for the subsequent content.

We utilize the model Eq.~(\ref{phase_inspiral}) to perform parameter estimation on observations of binary black holes obtained during the third observing run (O3) of the public LIGO-Virgo-KAGRA (LVK) data. The likelihood in the frequency domain assumes a colored Gaussian noise model, which is characterized by a power spectral density\cite{Talbot_19}. Our analysis incorporates uniform priors for the chirp mass, mass ratio, luminous distance, and relative deviation quadrupole moment $(\mathcal{M}, q, D_{L}, \Delta Q/Q)$. For each specific event, we generate the posterior probability density across the parameter space of these parameters.

Before conducting each parameter estimation, we execute a complete parameter estimation process to ensure consistency with the previous results obtained from the LVK. Subsequently, utilizing the same configuration, we initiate the computation of the posterior probability density across the parameter space of chirp mass, mass ratio, luminous distance, and relative deviation quadrupole moment to examine the validity of the no-hair theorem.

\begin{figure}
\centering
\includegraphics[width=0.50 \textwidth]{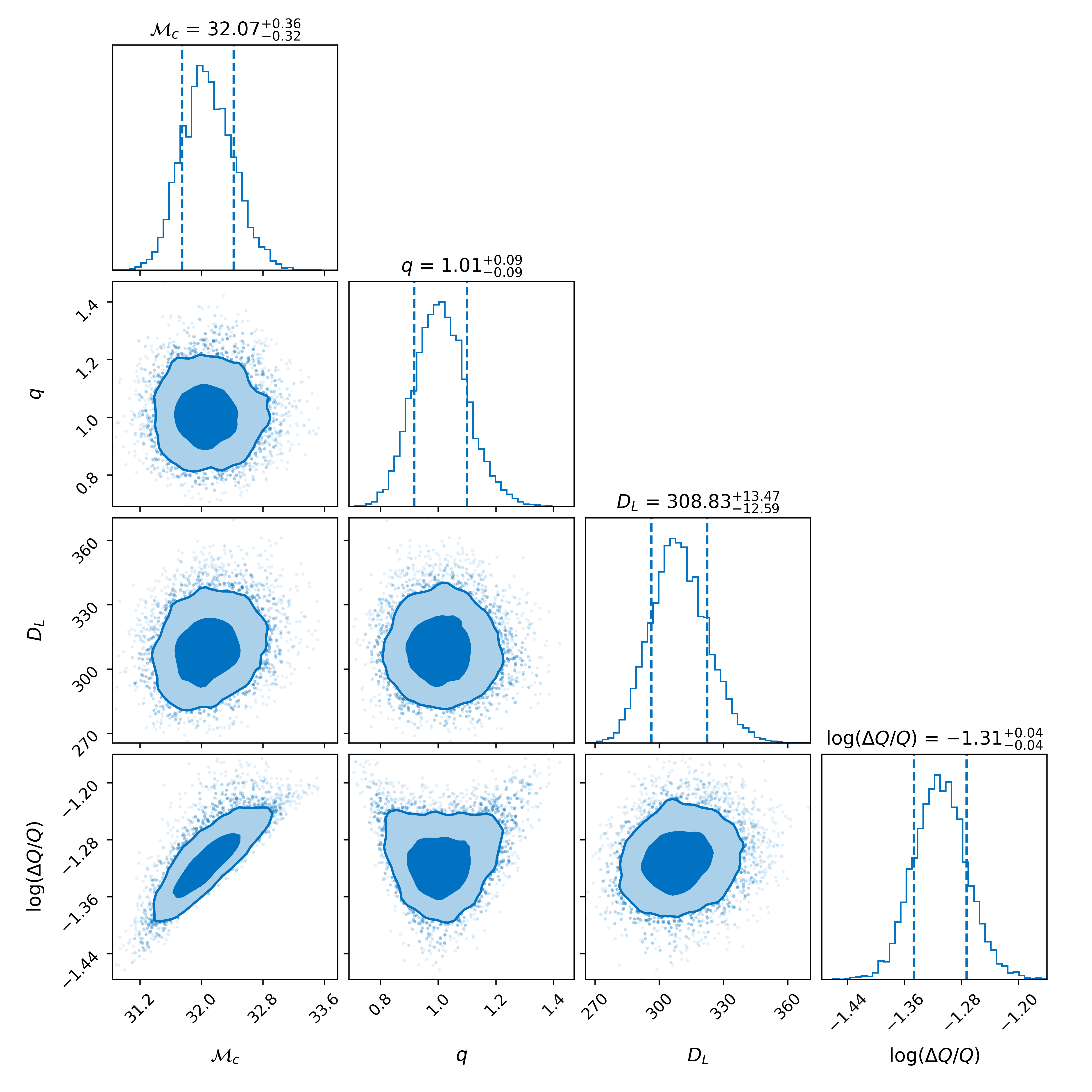}
\caption{Posterior for the parameters(chirp mass $\mathcal{M}_c$ , mass ratio $q$, luminous distance $D_L$, and relative deviation quadrupole moment $\Delta Q /Q$) for GW150914. The dashed blue lines in the 1D posteriors indicate the $1\sigma$ confidence levels of related parameters.}\label{posterior_GW150914}
\end{figure}

\textit{Results.}---Figure~\ref{posterior_GW150914} presents the posterior probability density for the chirp mass, mass ratio, luminous distance, and relative deviation in quadrupole moment for GW150914. The $1\sigma$ confidence levels of each related parameter are represented by dashed blue lines in each panel. Given the small value of the relative deviation quadrupole moment $\Delta Q/Q$, we employ the natural logarithm of $\Delta Q/Q$ (i.e., log($\Delta Q/Q$)) to display the results. The first three columns in Figure~\ref{posterior_GW150914} exhibit the posterior probability distribution of the chirp mass, mass ratio, and luminous distance, which aligns with the LVK's results. In the last column, we observe a non-zero value of the relative deviation quadrupole moment $\Delta Q/Q$, specifically 0.049. This implies that the non-general relativity full waveform model not only provides the corresponding quadrupole moment $Q_K$, but also a deviation $\Delta Q/Q$. The magnitude of $\Delta Q/Q$ indicates agreement with the no-hair theorem at approximately the $\sim$ 95\% level.

\begin{figure}
\centering
\includegraphics[width=0.50 \textwidth]{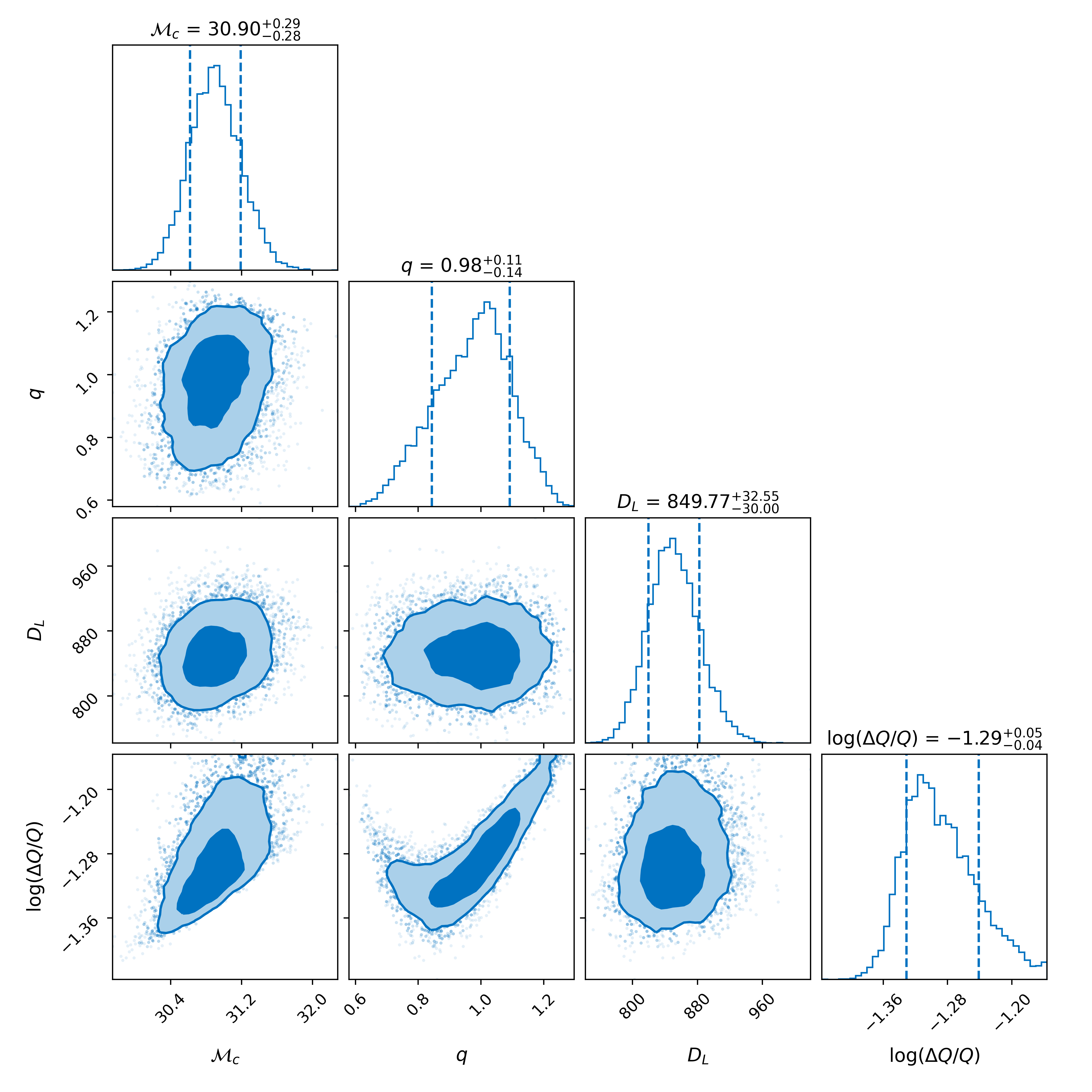}
\caption{Same as Fig.~\ref{posterior_GW150914}., but for GW200129}\label{posterior_GW200129}
\end{figure}

We conducted parameter estimation on observations of other binary black hole (BBH) mergers captured during the publicly available LIGO-Virgo-KAGRA (LVK) data. Our analysis reveals that the estimation of the relative deviation of the quadrupole moment $\Delta Q/Q$ is only possible for events with a high signal-to-noise ratio (SNR) and when the masses of the two black holes involved are comparable. In Table~\ref{Infor_GW_table}, we present four representative events. The first five columns provide the basic information about each event. Notably, the events GW200129\_065458 and GW150914 have both a high SNR and comparable mass, making it feasible to estimate the value of $\Delta Q/Q$. Referring to Fig.~\ref{posterior_GW150914} and \ref{posterior_GW200129} for the corresponding posterior distributions. Despite the high signal-to-noise ratio (SNR) of GW190814, we are unable to estimate the value of $\Delta Q/Q$ due to the incomparable masses of the two objects involved. Another potential reason is the uncertainty regarding whether the second compact object is a black hole, rendering our template inapplicable for this event. Conversely, GW190521 presents a different challenge, with comparable masses but a low SNR, which also precludes us from estimating $\Delta Q/Q$. Figure~\ref{posterior_GW200129} displays the posterior probability density for the same parameters as illustrated in Figure~\ref{posterior_GW150914}, and we can determine the relative deviation of the quadrupole moment for GW200129, yielding a value of $0.045$.

\begin{figure}
\centering
\includegraphics[width=0.50 \textwidth]{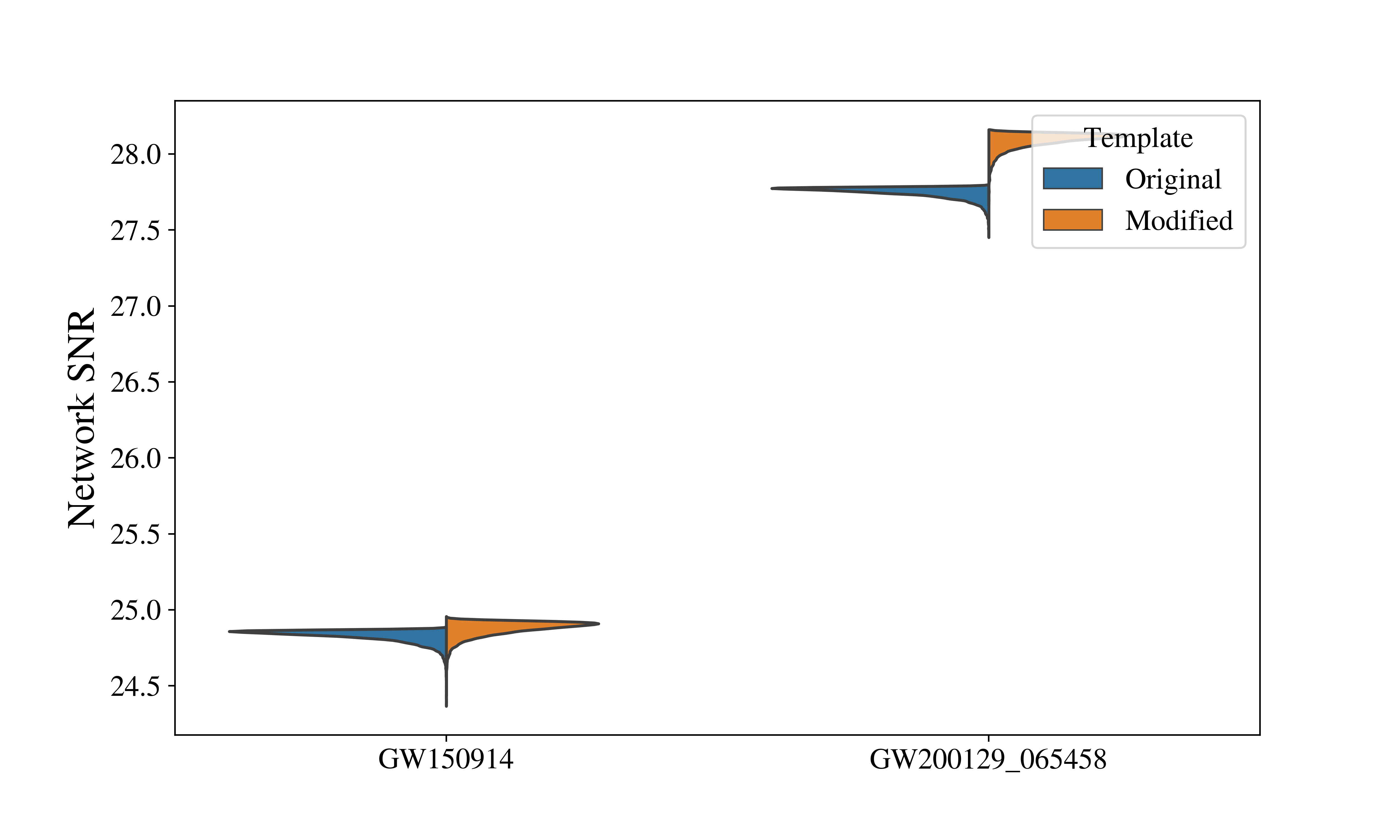}
\caption{The Network SNR for different templates where Original represents the IMRPhenomPv2 model and Modify represents the $\Psi_{\mathrm{FD}}$ model. The first column show the event GW150914 and the second one show GW200129\_065458.}\label{SNR_GW}
\end{figure}

Fig.~\ref{SNR_GW} presents the network signal-to-noise ratio (SNR) for different templates. The ``Original'' template corresponds to the IMRPhenomPv2 model, while the ``Modify'' template refers to the $\Psi_{\mathrm{FD}}$ model. The $\Psi_{\mathrm{FD}}$ model exhibits a slight increase in the network SNR compared to the IMRPhenomPv2 model. This may imply that our waveforms with deviated quadrupole moments are more suitable for the GW signals. For confirming this speculation, we compute the Bayes factor for the events GW150914 and GW200129, which are shown in the last column of Table.~\ref{Infor_GW_table}. It can be observed that the Bayes factor (log) for GW200129 is greater than the threshold 8, suggesting that the $\Psi_{\mathrm{FD}}$ model provides a better description for this event. Therefore the derivation from the Kerr black hole we found in GW200129 is statistically significant. However, one single example may not get the final conclusion, in the observations during O4 and future detections with space-based detectors, the availability of more data will enable more accurate tests to confirm or overturn our finding. However, the significant deviation from the Kerr scenario in the two events especially GW200129 should be given special attention.

\begin{figure}
\centering
\includegraphics[width=0.50 \textwidth]{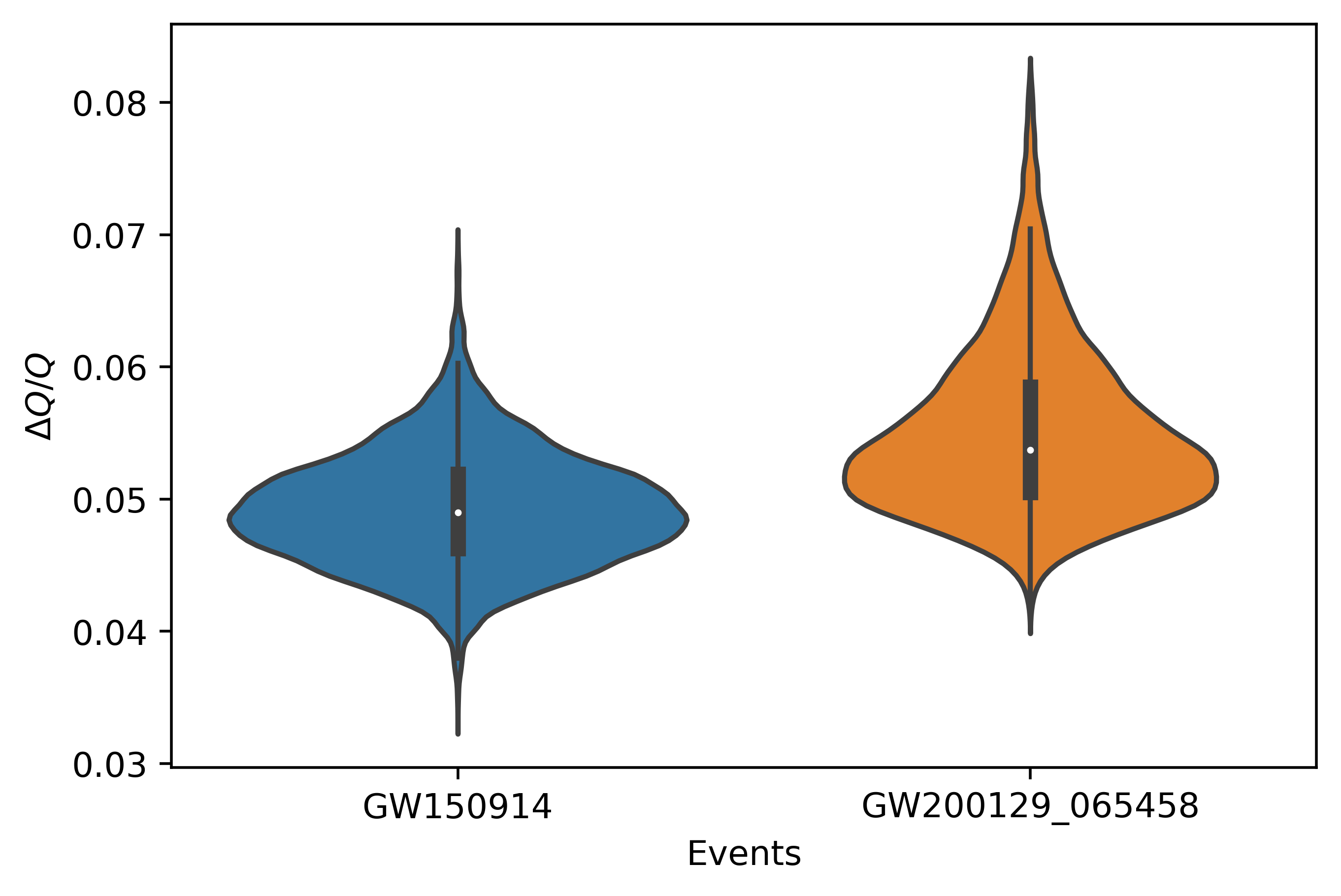}
\caption{The posterior probability density of relative deviation quadrupole moment $\Delta Q/Q$ for two events GW150914 and GW200129\_06548.}\label{posterior_Dou}
\end{figure}

To facilitate a comprehensive comparison of the relative deviation of the quadrupole moment between the two events, GW150914 and GW200129\_06548, we present the posterior probability density of the relative deviation quadrupole moment in Fig.~\ref{posterior_Dou}. In this figure, the white dot represents the median of the relative deviation quadrupole moment, while the black bar depicts the range of the quartile. Notably, both events exhibit a median relative deviation quadrupole moment of approximately 5\%, or on the other side, signify a consistent agreement with the no-hair theorem at the $\sim$ 95\% level. 

\begin{table*}[]
\renewcommand{\arraystretch}{1.3}
\begin{tabular}{|l|l|l|l|l|l|l|}
\hline
Event Name             & Mass1($M_{\odot}$) & Mass2($M_{\odot}$) & Chirp Mass($M_{\odot}$)  & Distance(Mpc) & Network SNR & Bayes factor(log) \\ \hline
GW200129\_065458 & $34.5^{+9.9}_{-3.2}$  & $28.9^{+3.4}_{-9.3}$  & $27.2^{+2.1}_{-2.3}$       & $900^{+290}_{-380}$    & $26.8^{+0.2}_{-0.2}$  & 10.61       \\ \hline 
GW150914         & $35.6^{+4.7}_{-3.1}$  & $30.6^{+3.0}_{-4.4}$  & $28.6^{+1.7}_{-1.5}$        & $440^{+150}_{-170}$           & $26.0^{+0.1}_{-0.2}$  &2.31     \\ \hline
\end{tabular}
\caption{The first column to the fifth column shows the median and 90\% symmetric credible intervals for selected source parameters from the LVK results. The last column shows the Bayes factor(log) for the different events, there are only two events, GW200129\_065458 and GW150914, that can estimate the Bayes factor.}
\label{Infor_GW_table}
\end{table*}

\textit{Discussion.}---The black hole in general relativity can be described by its mass, spin, and electric charge, as stated by the no-hair theorem. When the mass and spin are known, the quadrupole moment, represented by the value $Q=Ma^2$, can also be determined. Since the charge of most astrophysical black holes can be ignored, the quadrupole moment $Q$ does not contain any information about the charge. However, the deviation of the Kerr quadrupole moment may provide useful information if detected through gravitational waves, allowing for testing of the no-hair theorem. This relationship is demonstrated through the analysis of various gravitational events in this work.

We utilize the $\Psi_{\mathrm{FD}}$ waveform model to perform parameter estimation for gravitational events. The $\Psi_{\mathrm{FD}}$ waveform model is derived from the well-established IMRPhenom waveform models, and incorporates information about the deviation quadrupole moment, as indicated in Eq.~(\ref{phase_inspiral}). We use two events, GW150914 and GW200129\_06548 with high SNR and comparable mass-ratio, to estimate the deviation of quadrupole moment. In the O4 and future observations, more and more events with high SNRs will be found and should improve our tests of no-hair theorem based the $\Psi_{\mathrm{FD}}$ waveforms.  The results of parameter estimation are presented in Fig.~\ref{posterior_GW150914} and \ref{posterior_GW200129}. The estimated values for the chirp mass, mass ratio, and luminous distance are consistent with those obtained by the LVK results. Additionally, we determine the relative deviation of the quadrupole moment for GW150914 and GW200129\_06548 to be approximately 5\%, or in other words, provide a test of the no-hair theorem at a 95\% significance level. 

After calculating the network signal-to-noise ratio (SNR) of two events, we determined that GW150914 exhibited a similar value to the IMRPhenom waveform model, while GW200129\_06548 demonstrated a higher SNR when the $\Psi_{\mathrm{FD}}$ model was employed. Consequently, we computed the Bayes factor for GW200129\_06548 and obtained a value exceeding 8, indicating that the $\Psi_{\mathrm{FD}}$ model offers a more accurate depiction of this event. In this sense, we may carefully claim that we find a significant deviation from the Kerr black hole. However, this statement need to be confirm with more events with deeper analysis in the future. Figure.~\ref{posterior_Dou} displays the posterior probability density of the relative deviation quadrupole moment, revealing that GW150914 possesses a superior distribution of the relative deviation quadrupole moment, likely attributed to better estimation of the mass and luminous distance.

In the future, as more gravitational events are detected, particularly those with high SNRs, we will have the opportunity to investigate the no-hair theorem with greater precision. By studying these high SNR events, we can also explore the impact of intrinsic parameters on estimating the deviation quadrupole moment. This will provide insights into black holes' properties and offer a potential avenue for future study of general relativity. The 5\% deviation from the Kerr black hole we found in this Letter, though needs to be verified with more data and analysis, is still a significant milestone for the attempt of testing GR.

This work is supported by The National Key R\&D Program
of China (Grant No. 2021YFC2203002), NSFC (National Natural Science Foundation of China) No. 11773059 and No. 12173071. W. H. is supported by CAS Project for Young Scientists in Basic Research YSBR-006.

\bibliographystyle{apsrev4-1} 
%


\end{document}